# Kalinahia: Considering Quality of Service to Design and Execute Distributed Multimedia Applications

Sophie Laplace

Marc Dalmau

Philippe Roose

LIUPPA - IUT de Bayonne
Université de Pau et des Pays de l'Adour
Bayonne, France

Laplace@iutbayonne.univ-pau.fr

Marc.Dalmau@ieee.org

Roose@iutbayonne.univ-pau.fr

*Abstract*— One of the current challenges of Information Systems is to ensure semi-structured data transmission, such as multimedia data, in a distributed and pervasive environment. Information Sytems must then guarantee users a quality of service ensuring data accessibility whatever the hardware and network conditions may be. They must also guarantee information coherence and particularly intelligibility that imposes a personalization of the service. Within this framework, we propose a design method based on original models of multimedia applications and quality of service. We also define a supervision platform Kalinahia using a user centered heuristic allowing us to define at any moment which configuration of software components constitutes the best answers to users' wishes in terms of service.

## I. Introduction

One of the challenges of computing is to make more multimedia information and services available. The development of mobile devices and the widespread popularity of personal computers have created new needs for users wanting to have the same applications on their laptop and mobile peripherals. This ubiquity of applications induces strong variations of services and moreover discontinuities appear when the context of the physical support change. Simply providing a service is not enough, it is the quality of service which is essential for commercial success, particularly in the multimedia industry. Inevitably, people will not use mobile multimedia applications if they do not have a good quality of service – QoS. Indeed it is not possible to improve the context and also users need to have constant QoS so we chose to adapt applications to their context, whatever it is: hardware (network, etc.), environmental (brightness, etc.) or even end-user (special needs, languages, etc.). So, we propose to design an execution support allowing to adapt dynamically multimedia applications distributed on the Internet to the variations of their context in order to provide and maintain the best QoS possible to each individual user.

In this article, we will first compare our question to some relevant pieces of work. Then we will produce Kalinahia – Kalitatea Nahia, to seek quality in Basque language – a model of execution platform allowing us to optimize QoS and we present the most significant results obtained by our simulator.

## II. State of the Art

QoS was introduced initially in networks to describe the quality of the service provided by the communication systems to the applications. Then, its meaning extended [10] [19] from the purely technical aspects to concerns close to the user.

Thus this quality relies on the execution context of the application [16] [18] [1], in particular when this context varies in an unpredictable way. There are then two means of ensuring certain QoS to the user: to adapt the context to the application, or at least to guarantee the context, and to adapt the application to the context, including the user. However it is not always possible to work on the context especially as we include users. If we take into account the application adaptation from its design to its completion, the user will be able to get the best possible QoS.

We chose a field of study which seems most representative of these problems: distributed multimedia applications on Internet. Indeed, they are characterized [9] at the same time by high requirements for quality, a great sensitivity to the context and the strong variability of the context itself. The solutions usually suggested to maintain multimedia information systems with sufficient QoS in a variable context use either the resource allocation or a dynamic adaptation to the context [8].

Moreover, middleware set an effective tool to maintain certain QoS. For example, CAliF Multimedia [6] provides a middleware for cooperative multimedia applications by using the network resource allocation and the adaptation of the application needed to the available resources. The middleware Argilos [13] allows a hierarchical control of QoS by using not only the reservation of resources but also the configuration of components. JQoS [21] proposes an application of videoconference on Internet with adaptation of multimedia flow depending on the state of system QoS performances. Lastly, QuO [17] [20] allows QoS specification and management in applications built up with components: the platform and the application are adapted.

In the interest of making the user the focus of QoS concerns, we propose a middleware model enabling us to adapt dynamically the distributed multimedia applications to their execution context in order to maintain optimal QoS for users. This middleware adds or removes components and reconfigures the component assemblies.

## III. PLATFORM MODELING

We introduce Kalinahia a model of execution platform which deals in a distributed way with the deployment of the application as well as its supervision. We propose that the platform dynamically reconfigures the application by adapting its composition and its implementation. With this intention, each component of the operative part - components and flow - produced events of reconfiguration for the platform as soon as it detects a variation of its execution context. When a reconfiguration is necessary, the platform must propose a configuration offering better QoS. It is thus a question, "a priori", of finding an optimal assembly of components. However this problem is known to be NP-complete in the general case [7] [11].

### A. QoS and Application Models

We define QoS as the adequacy between the service expected by the user and the service provided so we model it more simply than in the general context of distributed applications [5] because only the characteristics of multimedia application which matter are the ones the user can directly perceive. We use two hierarchical levels [8], characteristics — simple QoS parameters — and criteria which gather characteristics according to their dependance — contextual criterion $Co$ — or not — intrinsic criterion $In$ — with respect to the context. We use a rating for each of the criteria to represent QoS of an entity: 0 for a crippling quality for the user and 1 for an optimal quality. So we chose as a model of QoS assessment the function which gives to QoS the value of the worst criterion like in the utility in microeconomics [15] [4].

>From the structural point of view, the application is built up according to the user's vision of the service provided - the Group. Each service is composed of various functionalities - the Sub-groups. The latter consist of software, hardware or human components, connected by information flows. The software components are encapsulated in Elementary Processors while the data flows are carried by Conducts [3].

In the general case of QoS evaluation, a characteristic is compared with the user's wishes so as to give a mark to it. The user will have attributed to the characteristics a relative weight which makes it possible to define Sub-Group QoS by a weighted average of the characteristics' marks. The QoS marks of application and Groups are obtained by similar averages.

### B. Algorithmic Complexity

Based on the fact that it was not possible to hope that the platform proposes the optimal configuration, we chose an approach taking into account the incidence of the reorganizations on the perception the user of the offered service has. Indeed, the user should not have to tolerate abrupt variations in the way the service is presented. Thus we proposed to implement a better configuration whose service is as close as possible to the configuration in the course of execution. We defined the proximity of service as follows: *Two configurations have nearby services where the user does not notice a change from one to the other.*

We can say that two configurations provide a close service if and only if their marks of the intrinsic and contextual criteria are close, which implies that the marks of the QoS are close. Because intrinsic criterion variation is more perceptible by users, platform will begin its research with the evaluation of the configurations having the same mark of intrinsic criterion as the current configuration. Thus it will initially only modify the mark of the contextual criterion.

If the new configuration is not optimal, the platform will be informed by new events of reconfiguration, which will enable to improve the QoS by a new research and thus to reach gradually the best configuration. Fig. 1 illustrates this principle of the iterative search for optimum.

The search for a better configuration is done by successively studying finite sets of configurations having close services called families. Each family provides a service of comparable nature and has the same mark of intrinsic criterion: thus they only differ by their adaptability to the context.

To be effective, the platform must target the modifications to be carried out on the application. In so far as each component of the operative part of the application is likely to generate events of reconfiguration, the information obtained by the platform is very precise and allows it to know which application components are problematic and which entities must be modified or removed. Thus, at first the platform will be able to restrict the scope of the study to the configurations, which differ from that in the course of execution only by the component at the origin of the reconfiguration event. However when this approach does not give any solution, we face the issue of the deployment ex-nihilo of a Sub-Group or a Group.

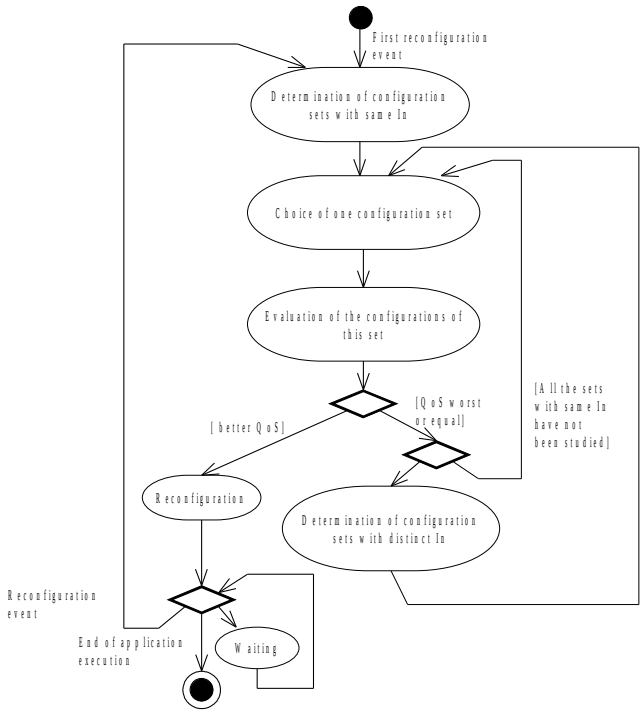

Figure1. Principle of the reconfigurations by the Kalinahia platform.

The reconfiguration events can result from passive or active measurements on the components and flows so as to detect the possible falls or improvements of QoS. They can also be transmitted by components called spies agents that the application designer introduced to collect non-measurable contextual information such as the language used during a video-conference [12]. For example, this information will be able to indicate that the service is no longer adapted for a listener who does not understand this language and then will impose a reconfiguration which offers a translation or subtitles.

*C. Kalinahia: Execution Platform Model*

We propose a platform model implementing iterative heuristics which improves the QoS with each iteration. The search for a better configuration rests then on two criteria. The first one is imposed by the temporal constraints of the multimedia applications. It is necessary that the platform quickly reacts in order to avoid service breaks. This is obtained by the generation of events of reconfiguration.

The second criterion also comes from the characteristics of the multimedia applications where the perception the user has of the application is central to evaluate QoS: it is the maintenance of ergonomic continuity at the time of a reconfiguration and it is called plasticity [2]. It is respected thanks to the study of the service proximity between configurations. Service proximity is determined, on the one hand, by using the architecture which we designed so that it reflects the vision the user of the service has and, on the other hand, by using the wishes the user will express.

Drawing from all these criteria, we built up a heuristic system. We have proved that its complexity is polynomial. It only depends on the intrinsic complexity of the application. In a logical way, this complexity is now incompressible.

## IV. KALINAHIA PLATFORM IMPLEMENTATION AND SIMULATION

*A. Implementation of the Platform*

The effectiveness of the heuristic that define reconfigurations depends partly on the choice of the event which identifies the problematic component. So we worked out an event manager model which allows the platform to intervene on the most critical application part for the user. Then we proposed a structural model of the platform allowing distributed management of the events but also the QoS evaluation with the aim of respecting the temporal constraints of multimedia applications. The management of the events is thus guided by their importance for the user. On all the stations used by the application, the local part of the platform is composed of five managers: the events manager carrying out the choice of the event to be dealt with, the evaluation manager, the communication manager, the user manager allowing the capture of users' wishes, the supervision manager reconfiguring the application.

*B. Simulation of the Platform*

We validated this model thanks to a simulator of the platform developed with Labview by National Instruments which makes it possible to simulate the application and its context of execution including the network.

The application is built using components which do not fulfill any function on the multimedia data. Their execution is simulated by the evolution of the QoS characteristics of outputs starting from the QoS characteristics of the inputs. These flows are represented by their QoS characteristics. The evaluation of QoS is then simulated by the definition of the QoS characteristics of the application output flows at a given time, as well as by their marking. The simulation of the application execution is obtained by continuous estimation of the QoS characteristics of all the flows present in the application. Our software makes it possible to simulate the state of the network and the stations used by the application. Thus, in the course of simulation, it is possible to vary the available bandwidth between two stations, the associated time of transmission and the saturation of a station.

The dynamic call of the components was carried out thanks to the use of a single model of component to program all the application components. The simulator creates, removes, and moves components in the same way a real platform would and with comparable times to a platform like OSGi [14]. According to the design of the application, to the users' wishes and to the information describing the context, the prototype simulates the operation of the application and the platform. It simulates the implementation of the local parts of the platform then dynamically displays the application by using the by default configurations. The application will then be carried out continuously until a reconfiguration is necessary. The prototype makes it possible at any moment to visualize the QoS mark of the application. In parallel, the platform collects the reconfiguration events. It identifies those of the highest priority and seeks, if necessary, a better configuration. If it fails in its search, it studies another event. If it succeeds, it sends an order of reconfiguration to the supervisor manager and the application is reconfigured.

We show here the results obtained for a video surveillance application [12] that offers 135 possible configurations: compression, picture processing, several qualities. One of the tests carried out (Fig. 2) characterize the performance of the platform and the application when the context of execution fluctuates. The application is first of all displayed in a favorable context – context 1, C.1 - where neither the stations nor the network are saturated then this context undergoes a degradation - context 2, C.2 – resulting from the saturation of the one of the stations. Then it goes back to its former state – C.1 – then gets deteriorated again – C.2.

The first reconfiguration corresponds to an improvement of the QoS and consists in replacing the component by a more powerful one. The second reconfiguration intervenes after a deterioration of the context and consists in moving a component from the saturated station to another station. The third reconfiguration improves the result obtained by moving another component to an unsaturated station.

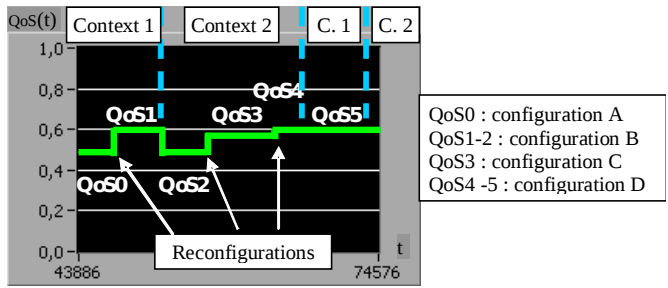

Figure2. Evolution of QoS at the time of context fluctuations

This test is particularly interesting because it highlights that, when the context goes back to its former situation after having moved, the platform proposes a QoS with identical mark but uses a different configuration that provides a satisfaction equivalent to the user. Moreover, by not seeking the optimum, the Kalinahia platform stabilizes the service and gives a better plasticity and thus a better performance to the application. An exhaustive research of the best configuration would not have this plasticity nor this stabilizing capacity.

## V. CONCLUSION

The platform that we propose carries out a particularly effective adaptation of the application to the context. Indeed the adaptation in Kalinahia is more dynamic than other works — CAliF, Agilos — because it is entirely dynamic since the principles of adaptation are determined in the course of execution according to the service proximity. Moreover, it relates at the same time to the structure, to the functionalities and to the scheduling, which makes it more complete than what is often proposed - JQoS, Agilos. It also automatically manages improvements and degradations of the context. Indeed, even if its objective at any moment remains to improve QoS as we defined it, in practice, this may imply a degradation of the intrinsic characteristics of the service so as to obtain an improvement of total quality thanks to a compromise between intrinsic and contextual criteria. So this ability to move in both directions of quality constitutes a progress compared to the systems usually suggested such as JQoS, where only degradation is carried out in an automatic way. Lastly, thanks to the heuristic system used, the platform does not just propose a solution to a NP-complete problem using only mathematical vision - as the solutions from graphs for networks - but also takes into consideration the user.

We think that platform and application must be closely dependent on each other, in order to leave the application with only the trade aspects, and thus to allow reuse of components or the use of components off-the-shelf. This is why we wish to produce not only an execution platform, but also an aid for the design. Lastly, studying the possibility of including our platform in a more ambitious system will be interesting, making it possible to use the adaptation we propose, but also to associate the exploitation of the guarantees of service when possible. Then we will be able to use an adaptation of the network regarding servers and using active nodes. Thus the service will be optimized from the user's point of view but also from the supplier's point of view, the original concern of the concept of QoS.

REFERENCES

[1] Ayed D., Taconet C., Bernard G., " A data model for context-aware deployment of component-based applications onto distributed systems", - ECOOP'04 - Oslo, Norway - 2004.

[2] Calvary G., Coutaz J., and Thevenin D., " A Unifying Reference Framework for the Development of Plastic User Interfaces " EHCI'01, Toronto, May 2001.

[3] Dalmau M., Roose P., Bouix E., Luthon F., " A Multimedia Oriented Component Model ", *IEEE 19th ICAINA*, Taiwan, March 2005.

[4] DaSilva L.A., " Pricing for QoS-enabled Networks: A Survey, " *IEEE Communication Surveys and Tutorials*, vol. 3, no. 2, 2000, pp. 2-8.

[5] Firesmith D.G., " Using Quality Models to Engineer Quality Requirements ", *JOT*, 2 (5), Swiss Federal Institute of Technology, Switzerland, p. 67-75, September/October 2003.

[6] Garcia E., Lapayre J.-C., Sureswaran R., and Tharmaraj K.. " Centralized or Distributed Algorithm for Concurrency Management in Multimedia Conferencing Systems ". *APAN 2001*, Penang, Malaysia, pages 108--119, August 2001.

[7] Garey M. R., Johnson D. S., *Computers and Interactability: A guide to the theory of NP-completeness*, W. H. Freeman and Company, San Francisco, 1979.

[8] Gu X., Nahrstedt K., Yuan W., Wichadakul D., XU D., " An XML-based QoS Enabling Language for the Web ", *Journal of Visual Language and Computing*, v.13, n.1, pp. 61-95, Academic Press, 2002.

[9] Hafid A., von Bochmann G., Dssouli R. "Distributed Multimedia Application and Quality of Service : A Review ", *Electronic Journal on Networks and Distributed Processing*, N°6, 1998, p 1-50.

[10] ITU, International Telecommunication Union, Rec.I.350, General Aspects of Quality of Service and Network Performance in Digital Networks, Geneva, 1989.

[11] Kuipers F., van Mieghem P., " MAMCRA: a constrained-based multicast routing algorithm ", *Proc. of Computer Communications*, vol. 25, pp.802-811, 2002.

[12] Laplace M. S., Conception d'architectures logicielles pour intégrer la qualité de service dans les applications multimédias réparties, Thèse de l'Université de Pau, 2006.

13] Li B., Kalter W., Nahrstedt K., " A Hierarchical Quality of Service Control Architecture for Configurable Multimedia Applications ", *Journal of High Speed Networks*, IOS Press, Vol. 9, pp. 153-174, 2001.
14] OSGi™ Alliance- The Dynamic Module System for Java - *About the OSGi ServicePlatform* - Technical Whitepaper, June 2007.
15] Parkin M., *Economics*, Addison-Wesley Inc., Reading, Massachusetts, 1990.
16] Rakotonirainy A., Indulska J., Loke S., Zaslavsky A., " Middleware for Reactive Components : An Integrated Use of Context, Roles, and Event Based Coordination ", *IFIP/ACM International Conference on Distributed Systems Platforms*, Heidelberg, Allemagne, Novembre 2001.
17] Schantz R.E., Loyall J.P., Rodrigues C., Schmidt D.C., Krishnamurthy Y., " Flexible and Adaptive QoS Control for Distributed Real-time and Embedded Middleware ". *ACM/IFIP/USENIX International Middleware Conf.*, Rio de Janeiro, Brazil, June 2003.
18] Stephen S., Yu W., Fariaz K., " Development of Situation-Aware Application Software for Ubiquitous Computing Environments ", in *Proc.of COMPSAC'02*, England, 2002.
19] Vogel A., Kerherve B., von Bochmann G., Gecei J., " Distributed multimedia applications and Quality of Service :- A Survey-" *IEEE Mutimedia Journal*, august 1995.
20] Wang N., Gill C., Schmidt D., Gokhale A., Natarajan B., Loyall J., Schantz R., Rodrigues C., " QoS-enabled Middleware ". Chap in *Middleware for Communications*, Qusay H. Mahmoud, Wiley, 2004.
21] Zhu W., Georganas N. " JQoS : Design and Implementation of a QoS-based Internet Videoconferencing System using Java Media Framework (JMF) " Canadian Conference on Electrical and Computer Engineering, Toronto, Canada, 2001.